\documentclass[showpacs,aps,graphicx,preprintnumbers,amsmath,twocolumn]{revtex4}%
\usepackage{amssymb}
\usepackage{graphicx}

\begin{document}
%\preprint{}
%\draft

\title{Photonic spatial Bell-state analysis for robust quantum secure direct communication using
quantum dot-cavity systems\footnote{Published in Eur. Phys. J. D
\textbf{67}, 30 (2013)}}

\author{Bao-Cang Ren, Hai-Rui Wei, Ming Hua, Tao Li,  and Fu-Guo Deng\footnote{Corresponding author. Email address:
fgdeng@bnu.edu.cn} }

\address{Department of Physics, Applied Optics Beijing Area Major Laboratory, Beijing Normal University, Beijing 100875, China }
\date{\today }

\begin{abstract}
Recently, experiments showed that the spatial-mode states of
entangled photons are more robust than their polarization-mode
states in quantum communications. Here we construct a complete and
deterministic protocol for  analyzing the spatial Bell states using
the interaction between a photon and an electron spin in a  charged
quantum dot inside a one-side micropillar microcavity. A quantum
nondemolition detector (QND) for checking the parity  of a
two-photon system can be constructed with the giant optical Faraday
rotation in this solid state system. With this parity-check QND, we
present a complete and deterministic proposal for the analysis of
the four spatial-mode Bell states. Moreover, we present a robust
two-step quantum secure direct communication protocol based on the
spatial-mode Bell states and the photonic spatial Bell-state
analysis. Our analysis shows that our BSA proposal  works in both
the strong and the weak coupling regimes if the side leakage and
cavity loss rate is small.
\end{abstract}


\maketitle

\section{Introduction}

Entanglement, an essential quantum correlation   between two quantum
subsystems \cite{book,Eng},  lies at the heart of quantum
information and it plays a critical role in quantum information
processing, especially for quantum computation \cite{book},  quantum
key distribution (QKD)  \cite{QKD1,QKD2,QKD3,QKD4,QKD5},  quantum
teleportation \cite{QT1},  entanglement swapping \cite{QS1}, quantum
dense coding \cite{DC1, DC2},  quantum secret sharing (QSS)
\cite{QSS,QSS2,QSS3}, and so on. In most quantum communication
tasks, Bell-state analysis (BSA) is required, especially  in
teleportation,  entanglement swapping, dense coding, quantum state
sharing \cite{QSTS1,QSTS2,QSTS3,QSTS4,QSTS5},  and quantum
repeaters.  BSA is defined as the projection of the state of a
two-qubit system onto the maximally entangled Bell states. The four
Bell states of two-qubit systems are orthogonal to each other and
are  deemed to be discriminated deterministically in principle, but
BSA is always a technical difficulty, which yields a low efficiency
and a low discrimination fidelity \cite{lixhijqi}. It is impossible
to distinguish the four Bell states in the polarization degree of
freedom (DOF)  with only linear optical elements.  In 1999,
Vaidman's \cite{BSA1} and L$\ddot{u}$tkenhaus' \cite{BSA2} groups
put forward a  BSA protocol for teleportation with the success
probability of $\frac{1}{2}$, resorting to only linear optical
elements.  In the following years, some other proposals to
 analyze completely Bell states in the  polarization DOF, including
those using spatial entanglement of photons, were introduced
\cite{BSAT,BSAE1,BSAE2,BSAE3,BSA1kwiat,BSA2walborn,BSA3,BSA4}. The
complete BSA on the  polarization DOF requires the hyperentanglement
in both the polarization DOF and the spatial DOF (or another DOF) or
the nonlinear interaction between photons and media.

Recent works showed that an electron spin in a charged quantum dot
(QD) inside a microwave cavity \cite{3} can be used to store and
process quantum information.  In 2008,  Hu  et al.  \cite{QD1}
proposed a quantum nondemolition method using the interaction of the
left-circularly and the right-circularly polarized lights with a
QD-cavity system. This giant circular birefringence (GCB) in a
QD-cavity system can be used to construct a CNOT gate \cite{QD2}, a
multiphoton entangler \cite{QD1,QD3,QD4},  a photonic polarization
Bell-state analyzer \cite{QD5}, entanglement purification
\cite{QD6}, and so on.

In 2000, Long and Liu \cite{LongLiu} proposed an interesting quantum
communication protocol with a quantum data block, which can be used
to transmit the secret message  directly between two legitimate
users (without creating a private key first). It was detailed and
given the term as quantum secure direct communication (QSDC) by
Deng, Long, and Liu \cite{two-step} in 2003. Subsequently, QSDC was
actively pursued by some groups and people proposed some QSDC
protocols
\cite{LongLiu,two-step,QOTP,highdimension,lixhcp,dengpla,dengchinphys,wangtjcpl,Gub1,Liud,Sunzw,Shij,Gaog,Yangc},
quasi-secure  quantum communication protocols \cite{Pingpong} and
some deterministic secure quantum communication protocols
\cite{lixhjkps,Yan1,Manzhang,Zhuxia,LeeLim,WangZhang,Gaot1}. In most
of QSDC protocols existing, the polarization states of photon
systems are used to carry the secret message \cite{QSDCreview}.
However, the polarization states of photons are fragile to channel
noise when they are transmitted over a practical channel, such as an
optical fiber or a free space. Fortunately, some DOFs of photon
systems are robust against channel noise,  such as the spatial-mode
DOF and the frequency DOF of photon systems. For example,  the
entanglement in the spatial-mode DOF of photon pairs was used to
purify the entanglement in the polarization DOF by people
\cite{eppsimon,eppshengpdc,eppshengdeter,epplixhdeter,eppshengone,eppdengone}.
The experiment showed \cite{epppan} that it is not more difficult to
control the stability of the spatial entanglement. In 2008, the
experiment by Min$\acute{a}$$\breve{r}$  et al.  \cite{phaseexp} for
phase-noise measurements showed that the phase in long fibers with
several tens of km remains stable in a realistic environment, which
is an acceptable level for time on the order of 100 $\mu$s and is
enough for the phase stabilization in feasible quantum repeaters
with optical-fiber networks.  Also,  the frequency entanglement is
used for faithful entanglement distribution on the polarization
entanglement of photon pairs \cite{distrisheng}. In 1999,  Merolla
et al.  \cite{frequencyexp}  showed that a quantum bit error rate
contribution of approximately 4\% and estimated that 2\% could be
attributed to the transmission of the central frequency by the
Fabry-Perot cavity in a key distribution experiment over a 20-km
single-mode  optical-fiber  spool \cite{eppshengdeter}. However, the
Hadamard operation on the frequency  states of photons is difficult,
which limits the direct application of frequency entanglement in
quantum communication, different from the spatial entanglement of
photon systems.

In this paper,  we present a complete and deterministic analysis
protocol for the photonic spatial Bell states, resorting to the GCB
based on a quantum dot inside a one-side micropillar microcavity. In
our proposal, we first construct  a parity-check quantum
nondemolition detector (QND) for the spatial-mode DOF of photon
pairs by exploiting the interaction between a photon and an electron
spin in a charged quantum dot (QD) inside a one-side micropillar
microcavity. With this QND, we can first divide the four Bell states
in spatial-mode DOF into two groups, that is, the even-parity group
and the odd-parity group, according to their parities of the two
photons in the spatial-mode DOF, and then distinguish the relative
phase of the two states in each group by transforming the difference
of the phases into that of the spatial modes with BSs and QND. Based
on the spatial-mode Bell states and their BSA, we present a robust
two-step quantum secure direct communication protocol as the spatial
entanglement of two-photon systems is more robust than the
polarization entanglement. Our analysis shows that our BSA proposal
for spatial-mode Bell states works in both the strong coupling
regime and the weak coupling regime  if the side leakage and cavity
loss rate is small.

%section 2

\section{QD-cavity system}

In order to understand the principle of the interaction between a
photon and an electron spin, we first introduce the construction of
a one-side QD-cavity system. The QD-cavity system is always
constructed by a singly charged QD (self-assembled In(Ga)As QD or a
GaAs interface QD) located in the center of a one-side optical
resonant cavity for maximal light-matter coupling. If an excess
electron is injected into the singly charged QD,  a negatively
charged exciton ($X^-$) with two electrons bound to one hole
\cite{QD7} is created by optical excitation. Different excess
electron spins  create different structures of  $X^-$, according to
Pauli's  exclusion principle, which is called $X^-$ spin-dependent
transitions \cite{QD8}.  If the injected excess electron spin is in
the state $|\uparrow\rangle$,  the QD-cavity system resonantly
absorbs a left-handed circularly polarized light $|L\rangle$ and
creates a negatively charged exciton in the state
$|\uparrow\downarrow\Uparrow\rangle$. If the injected excess
electron spin is in the state $|\downarrow\rangle$, the QD-cavity
system resonantly absorbs a right-handed circularly polarized light
$|R\rangle$ and creates a negatively charged exciton in the state
$|\downarrow\uparrow\Downarrow\rangle$.  Here,  $|\Uparrow\rangle$
and $|\Downarrow\rangle$ represent the heavy-hole spin states
$|+\frac{3}{2}\rangle$ and $|-\frac{3}{2}\rangle$, respectively.

The $X^-$ spin-dependent transitions have different affections on
the phase shifts of two circularly polarized lights after being
reflected from the optical resonant microcavity. This optical
transition process can be characterized by Heisenberg equations for
the cavity-field operator $\hat{a}$ and $X^-$ dipole operator
$\sigma_-$ in the interaction picture \cite{QD9} as follows:
\begin{eqnarray}                           \label{eq.1}
\frac{d\hat{a}}{dt}  & = &  -[i(\omega_c-\omega)+\frac{\kappa}{2}+\frac{\kappa_s}{2}]\,\hat{a}-g\sigma_{-} - \sqrt{\kappa}\,\hat{a}_{in}, \nonumber\\
\frac{d\sigma_-}{dt} & = &  -[i(\omega_{X^-}-\omega) + \frac{\gamma}{2}]\,\sigma_--g\sigma_z\,\hat{a},\nonumber\\
\hat{a}_{out}        & = &  \hat{a}_{in} + \sqrt{\kappa}\,\hat{a},
\end{eqnarray}
where $g$ describes the coupling strength of $X^-$ and the cavity
mode. $\omega_c$,  $\omega$ and $\omega_{X^-}$  describe the
frequencies of  the cavity mode, the input probe light and the $X^-$
transition, respectively. $\gamma/2$ describes the decay rates of
$X^-$. $\kappa/2$ and $\kappa_s/2$ describe the decay rates and the
side leakage rate of the cavity respectively.

\begin{figure}[!h]%[tpb]        %Figure 1
\centering\includegraphics[width=8.0 cm]{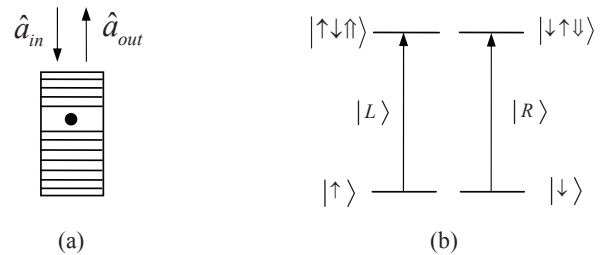} \caption{The
$X^-$ spin-dependent transitions. (a) A singly charged QD inside a
one-side micropillar microcavity with a circular cross section. (b)
$X^-$ spin selection transition rule due to the Pauli's exclusion
principle. $L$ and $R$  describe  the left and the right circularly
polarized lights, respectively. } \label{figure1}
\end{figure}

Considering the weak excitation condition with $X^-$ staying in the
ground state at most time ($\langle\sigma_z\rangle=-1$),  the
reflection coefficient of the QD-cavity system is \cite{QD1}
\begin{eqnarray}                           \label{eq.2}
r(\omega)=1-\frac{\kappa[i(\omega_{X^-}-\omega)+\frac{\gamma}{2}]}{[i(\omega_{X^-}-\omega)
+\frac{\gamma}{2}][i(\omega_c-\omega)+\frac{\kappa}{2}+\frac{\kappa_s}{2}]+g^2}.\nonumber\\
\end{eqnarray}
If the $X^-$ and the cavity mode are uncoupled (a cold cavity), the
coupling strength is $g=0$ and the reflection coefficient
$r_0(\omega)$ is \cite{QD1}
\begin{eqnarray}                           \label{eq.3}
r_0(\omega)=\frac{i(\omega_c-\omega)-\frac{\kappa}{2}+\frac{\kappa_s}{2}}{i(\omega_c-\omega)+\frac{\kappa}{2}+\frac{\kappa_s}{2}}.
\end{eqnarray}

Considering $X^-$ spin-dependent transitions, the phase shift rules
of two circularly polarized lights can be obtained. If the excess
electron spin is in the state $|\uparrow\rangle$,  the $|L\rangle$
light gets a reflection phase shift $\varphi_h$ with a hot cavity
($X^-$ and the cavity mode are coupled), and the  $|R\rangle$ light
gets a reflection phase shift $\varphi_0$ with a cold cavity. If the
excess electron spin is in the state $|\downarrow\rangle$,  the
$|L\rangle$ light gets a reflection phase shift $\varphi_0$ with a
cold cavity, and the $|R\rangle$ light gets a reflection phase shift
$\varphi_h$ with a hot cavity. If one adjusts the frequencies
$\omega$ and $\omega_c$, the reflection coefficient can reach
$|r_0(\omega)|\cong1$ and $|r_h(\omega)|\cong1$ in an ideal
condition. With the initial states of an electron spin and a
circularly polarized light in the superposition states
$\frac{1}{\sqrt{2}}(|\uparrow\rangle+|\downarrow\rangle)$ and
$\frac{1}{\sqrt{2}}(|L\rangle+|R\rangle)$, the two systems become
entangled after the light is reflected by the cavity, that is,
\begin{eqnarray}                           \label{eq.4}
&&\frac{1}{\sqrt{2}}(|R\rangle+|L\rangle)\otimes \frac{1}{\sqrt{2}}(|\uparrow\rangle+|\downarrow\rangle)\nonumber\\
&&\;\;\;\;\rightarrow
 \frac{1}{2}e^{i\varphi_0}[(|R\rangle+e^{i\Delta\varphi}|L\rangle)|\uparrow\rangle
+(e^{i\Delta\varphi}|R\rangle+|L\rangle)|\downarrow\rangle],\nonumber\\
\end{eqnarray}
where $\Delta\varphi=\varphi_h-\varphi_0$,
$\varphi_0=arg[r_0(\omega)]$ and $\varphi_h=arg[r_h(\omega)]$.
$\theta_F^\uparrow=(\varphi_0-\varphi_h)/2=-\theta_F^\downarrow$ is
the Faraday rotation angle.

\section{Bell-state analysis of spatial-mode entanglement}

% section 3

\begin{figure}[!h]%[tpb]        %Figure 2
\centering\includegraphics[width=7 cm,angle=0]{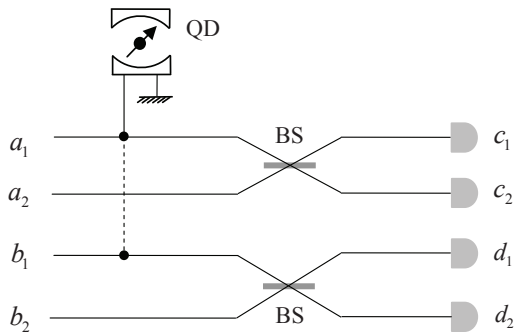}
\caption{Schematic  diagram for analyzing the spatial-mode Bell
states completely. The QND is used to distinguish the odd-parity
states $|\psi^\pm\rangle$ from the even-parity states
$|\phi^\pm\rangle$. BS represents a 50:50 beam splitter.}
\label{figure2}
\end{figure}

The photonic spatial Bell-states usually have the form as follows
\begin{eqnarray}                           \label{eq.5}
|\phi^\pm\rangle_{ab} &=& \frac{1}{\sqrt2}(|a_1b_1\rangle\pm|a_2b_2\rangle)_{ab},\nonumber\\
|\psi^\pm\rangle_{ab} &=&
\frac{1}{\sqrt2}(|a_1b_2\rangle\pm|a_2b_1\rangle)_{ab},
\end{eqnarray}
where $a$ and $b$ present two photons, and $a_1$ ($b_1$) and $a_2$
($b_2$) are the different spatial modes for the photon $a$ ($b$),
shown in Fig.\ref{figure2}. The states $|\phi^\pm\rangle$ are always
called even-parity states and the  states $|\psi^\pm\rangle$ are
always called odd-parity states.

To distinguish these four Bell states in the spatial-mode DOF, we
should first divide the four Bell states into two groups, that is,
even-parity states and odd-parity states. This task can be
accomplished with  a parity-check quantum nondemolition detector
(QND). This parity-check QND can be constructed with a one-side
QD-cavity system, shown in  Fig.\ref{figure2}.  Let us assume that
the excess electron spin is initially prepared in the state
$\frac{1}{\sqrt{2}}(|\uparrow\rangle+|\downarrow\rangle)_e$ and the
spatial Bell state of the two-photon system $ab$ is  in  the
superposition of two circularly polarized components
$(\alpha|R\rangle + \beta|L\rangle)_a(\alpha'|R\rangle +
\beta'|L\rangle)_b$ (here
$|\alpha|^2+|\beta|^2=|\alpha'|^2+|\beta'|^2=1$). If the frequencies
of the input photon and the cavity mode are adjusted as
$\omega-\omega_c\approx\kappa/2$ to get $\Delta\varphi=\pi/2$, after
a photon interacts with the QD-cavity twice, the state of the system
composed of the input photon and the electron spin becomes
\begin{eqnarray}                           \label{eq.6}
(\alpha|R\rangle+\beta|L\rangle)_a\otimes \frac{1}{\sqrt{2}}(|\uparrow\rangle+|\downarrow\rangle)_e \rightarrow(\alpha|R\rangle-\beta|L\rangle)_a\nonumber\\
\otimes  \frac{1}{\sqrt{2}}(|\uparrow\rangle-|\downarrow\rangle)_e.
\end{eqnarray}
By measuring the excess electron spin in QD with the orthogonal
basis
$\{|\pm\rangle=\frac{1}{\sqrt{2}}(|\uparrow\rangle\pm|\downarrow\rangle)\}$,
one can see  that the state of the excess electron spin changes if
there is a photon interacting with the QD-cavity system (with a
phase-flip operation $U_P=\vert R\rangle\langle R\vert - \vert
L\rangle\langle L\vert$ on the photon $a$, its original state is
recovered). If there are two photons interacting with the QD-cavity
system, the state of the electron spin becomes unchanged again.

In order to measure the excess electron spin, the orthogonal basis
$\{|\pm\rangle\}$ is first rotated to the orthogonal basis
$\{|\uparrow\rangle, |\downarrow\rangle\}$ by applying a Hadamard
gate. With an auxiliary photon $p$, whose original state is
$\frac{1}{\sqrt2}(|R\rangle + |L\rangle)_p$, put into the cavity,
where the photon only interacts with the QD-cavity system once, the
state of the system composed of the auxiliary photon and the
electron spin in the QD after reflection becomes
\begin{eqnarray}                           \label{eq.7}
 \frac{1}{\sqrt{2}}(|R\rangle+|L\rangle)_p|\uparrow\rangle   &\rightarrow&  \frac{1}{\sqrt{2}}(|R\rangle+i|L\rangle)_p|\uparrow\rangle,\nonumber\\
 \frac{1}{\sqrt{2}}(|R\rangle+|L\rangle)_p|\downarrow\rangle &\rightarrow&  \frac{1}{\sqrt{2}}(|R\rangle-i|L\rangle)_p|\downarrow\rangle.
\end{eqnarray}
By measuring the state of the output auxiliary photon $p$ in the
orthogonal linear polarizations $\frac{1}{\sqrt{2}}(|R\rangle \pm
i|L\rangle)$, one can distinguish whether the state of the excess
electron spin in the QD is changed or not. In detail, if
 the auxiliary photon is  in the state $\frac{1}{\sqrt{2}}(|R\rangle +
i|L\rangle)$, the state of the electron spin $e$ does not change.
However, if the  auxiliary photon is  in the state
$\frac{1}{\sqrt{2}}(|R\rangle - i|L\rangle)$, the state of the
electron spin $e$ is  changed.

By far,  we have constructed a photon-number QND to distinguish an
even number of photons from an odd number of photons. If the excess
electron spin is changed, there are an odd number of photons passing
through the QD-cavity system. If the excess electron spin is
unchanged, there are an even number of photons passing through the
QD-cavity system. In Fig.\ref{figure2},  we construct a parity-check
QND with this photon number QND. If the state of the excess electron
spin in QD is changed, the spatial Bell state input is one of the
two odd-parity states $|\psi^\pm\rangle$. The input spatial Bell
state is one of the two even-parity states $|\phi^\pm\rangle$ if the
state of the excess electron spin in QD is unchanged.

After dividing the spatial Bell states into an odd-parity states
group and  an even-parity states group, we have to distinguish the
different relative phases in each group. In order to distinguish the
Bell state with a relative phase zero from the Bell state with a
relative phase  $\pi$, Hadamard operations are performed on the two
photons with two 50:50 beam splitters (BS) shown in
Fig.\ref{figure2}. A BS can be used to complete the transformation
of photon states in the spatial-mode DOF:
\begin{eqnarray}                           \label{eq.8}
|a_1\rangle &\rightarrow& \frac{1}{\sqrt2}(|c_1\rangle+|c_2\rangle),\nonumber\\
|a_2\rangle &\rightarrow& \frac{1}{\sqrt2}(|c_1\rangle-|c_2\rangle),\nonumber\\
|b_1\rangle &\rightarrow& \frac{1}{\sqrt2}(|d_1\rangle+|d_2\rangle),\nonumber\\
|b_2\rangle &\rightarrow& \frac{1}{\sqrt2}(|d_1\rangle-|d_2\rangle).
\end{eqnarray}
The two groups are transformed as follows:
\begin{eqnarray}                           \label{eq.9}
\frac{1}{\sqrt2}(|a_1b_1\rangle+|a_2b_2\rangle) &\rightarrow& \frac{1}{\sqrt2}(|c_1d_1\rangle+|c_2d_2\rangle), \nonumber\\
\frac{1}{\sqrt2}(|a_1b_1\rangle-|a_2b_2\rangle) &\rightarrow& \frac{1}{\sqrt2}(|c_1d_2\rangle+|c_2d_1\rangle),\nonumber\\
\frac{1}{\sqrt2}(|a_1b_2\rangle+|a_2b_1\rangle) &\rightarrow& \frac{1}{\sqrt2}(|c_1d_1\rangle-|c_2d_2\rangle),\nonumber\\
\frac{1}{\sqrt2}(|a_1b_2\rangle-|a_2b_1\rangle) &\rightarrow&
\frac{1}{\sqrt2}(|c_1d_2\rangle-|c_2d_1\rangle).
\end{eqnarray}
This means that the states $|\phi^+\rangle_{ab}$,
$|\phi^-\rangle_{ab}$, $|\psi^+\rangle_{ab}$ and
$|\psi^-\rangle_{ab}$ will become $|\phi^+\rangle_{cd}$,
$|\psi^+\rangle_{cd}$, $|\phi^-\rangle_{cd}$ and
$|\psi^-\rangle_{cd}$, respectively. With the four detectors shown
in Fig.\ref{figure2}, one can completely distinguish the four Bell
states in the spatial-mode DOF. If the detectors $c_1$ and  $d_1$ or
$c_2$ and $ d_2$ click, the input state is $|\phi^+\rangle_{ab}$ or
$|\psi^+\rangle_{ab}$, respectively. If the detectors $c_1$ and
$d_2$ or $c_2$ and  $d_1$ click, the input state is
$|\phi^-\rangle_{ab}$ or $|\psi^-\rangle_{ab}$, respectively.

Now we have completely distinguish the four spatial-mode Bell
states. The relation of the input Bell states and the measurement of
the excess electron spin and detectors clicked  is shown in Table
\ref{tab1}. If the excess electron spin in QD is unchanged and the
detectors $c_1, d_1$ or $c_2, d_2$ click, the state input is
$|\phi^+\rangle$. If the excess electron spin in QD is changed and
the detectors $c_1$ and $d_1$ or $c_2$  and $d_2$ click, the input
state is $|\psi^+\rangle$. The excess electron spin in QD is
unchanged and the detectors $c_1$ and $d_2$ or $c_2$ and $d_1$ click
for the input state $|\phi^-\rangle$, and the excess electron spin
in QD is changed and the detectors $c_1$ and $d_2$ or $c_2$ and
$d_1$ click for the input state $|\psi^-\rangle$.

\begin{table}
\caption{Output results for complete spital Bell-state analysis.}
\begin{ruledtabular}
\begin{tabular}{ccc}

      & \multicolumn {2}{c}{Results} \\  \cline{2-3}
    State               &          QD           &            Detectors                  \\
   \hline

  $|\psi^{+}\rangle$    &       change          &      $c_1, d_1$ or $c_2, d_2$               \\

  $|\psi^{-}\rangle$    &       change          &      $c_1, d_2$ or $c_2, d_1$                 \\

  $|\phi^{+}\rangle$    &       unchange        &      $c_1, d_1$ or $c_2, d_2$                \\

  $|\phi^{-}\rangle$    &       unchange        &      $c_1, d_2$ or $c_2, d_1$              \\
\end{tabular}\label{tab1}
\end{ruledtabular}
\end{table}

\section{Robust two-step QSDC based on spatial entanglement}

We will present a robust two-step QSDC protocol, following the ideas
in Refs.\cite{LongLiu,two-step}. With photon pairs in a spatial-mode
Bell state, the principle of this robust two-step QSDC can be
described in detail as follows.

(1) The phase for creating a secure quantum channel between the two
users with photon pairs in a spatial-mode Bell state.

In this time, Alice and Bob can create a secure quantum channel with
a sequence of photon pairs in a spatial-mode Bell state. In detail,
Bob prepares a sequence of photon pairs in which each pair is in the
spatial-mode Bell state $\vert
\phi^+\rangle_{ab}=\frac{1}{\sqrt{2}}(\vert a_1b_1\rangle + \vert
a_2b_2\rangle)_{ab}$. Bob divides the photon pairs into two photon
sequences $S_A$ and $S_B$. That is, $S_A$ ($S_B$) is composed of all
the photons $a$ ($b$) in the photon pairs, as the same as that in
the original two-step protocol \cite{LongLiu,two-step}. Bob sends
the photon sequence $S_A$ to Alice. After Alice receives the
sequence $S_A$, she chooses randomly a subset of photon samples from
$S_A$ and measures them by choosing randomly one of two
nonorthogonal bases for each photon, say $Z$ and $X$. Here
$Z=\{\vert a_1\rangle, \vert a_2\rangle\}$ and $X\equiv \{\vert \pm
\rangle=\frac{1}{\sqrt{2}}(\vert a_1\rangle \pm \vert
a_2\rangle)\}$. She tells Bob the positions and the outcomes of
sampling photons. Bob measures the photons correlated to the samples
chosen by Alice with the same bases as those by Alice and then
analyzes the security of their quantum channel with a sequence of
photon pairs in a spatial-mode Bell state. If Bob confirms that
their channel is secure, he tells Alice to encode her secret message
on the photons remained in the photon sequence $S_A$. Otherwise, Bob
and Alice discard their quantum channel and repeat their quantum
communication from the beginning.

In essence, Bob and Alice create a secure quantum channel with a
sequence of photon pairs in a spatial-mode Bell state in the first
phase. The eavesdropping check can be accomplished with the
correlation between the two photons in each pair, as the same as the
Bennett-Brassard-Mermin QKD protocol \cite{QKD2}. More accurately,
it is as the same as the eavesdropping check in the original
two-step QSDC protocol  \cite{LongLiu,two-step}  in principle.
However, this process is more robust than that in the original
two-step QSDC protocol  \cite{LongLiu,two-step} as the spatial-mode
entanglement is more stable than the polarization entanglement of
photon pairs, as shown in previous works
\cite{eppsimon,eppshengpdc,eppshengdeter,epplixhdeter,eppshengone,eppdengone,epppan}.

(2) The message coding and decoding phase.

After Alice and Bob create a quantum channel securely, they can
exchange the secret message directly. That is, Alice encodes her
secret message on the photons in the sequence $S_A$ with four
unitary operations $U_i$ ($i=1,2,3,4$) in the spatial-mode DOF. Here
\begin{eqnarray}                           \label{eq.10}
U_1 &=& \vert a_1\rangle\langle a_1\vert +  \vert a_2\rangle\langle a_2\vert, \nonumber\\
U_2 &=& \vert a_1\rangle\langle a_2\vert +  \vert a_2\rangle\langle a_1\vert, \nonumber\\
U_3 &=& \vert a_1\rangle\langle a_1\vert -  \vert a_2\rangle\langle a_2\vert, \nonumber\\
U_4 &=& \vert a_1\rangle\langle a_2\vert -  \vert a_2\rangle\langle
a_1\vert.
\end{eqnarray}
With these unitary operations, Alice can transform the spatial-mode
Bell state $\vert \phi^+\rangle_{ab}$ into one of the four Bell
states $\vert \phi^\pm\rangle_{ab}$ and $\vert
\psi^\pm\rangle_{ab}$.

In order to check eavesdropping, Alice chooses randomly some photons
in the sequence $S_A$ as the samples for error rate analysis and she
encodes them with the four unitary operations randomly. After coding
the message, Alice sends the photon sequence $S_A$ to Bob. By
combining the two photon sequences $S_A$ and $S_B$ and performing a
BSA on the spatial-mode entangled states of each photon pair, Bob
can obtain the information about the unitary operations chosen by
Alice in a deterministic way. After filtering out the sample pairs
for eavesdropping check, Bob can, in principle, read out the secret
message sent by Alice directly if they code the operations $U_1$,
$U_2$, $U_3$, and $U_4$ as the message 00, 01, 10, and 11,
respectively.

From the description above, one can see that this robust two-step
QSDC protocol does not increase the difficulty of its implementation
in experiment, compared with the original two-step QSDC protocol
\cite{LongLiu,two-step}. Except for the complete analysis for the
four spatial-mode Bell states, all the unitary operations can be
accomplished with linear optical elements, such as BSs. However, it
improves the robustness of the two-step QSDC protocol largely.
Moreover, the present robust two-step QSDC protocol requires Alice
to send the photon sequence $S_A$ back to Bob, not the case that Bob
sends both the sequences $S_B$ and $S_A$ to Alice, which is
different to the original two-step QSDC protocol
\cite{LongLiu,two-step}. This modification can reduce the
requirement of entanglement swapping when there is photon loss in
the transmission line. Certainly, Alice and Bob should prevent a
vicious eavesdropper from stealing their secret message with Trojan
horse attack \cite{lixhattack,dengattack}.

\section{Discussion and summary}

Bell states analysis is necessary in quantum communication. Many
works were focused on a complete and deterministic analysis of Bell
states in the  polarization DOF of photon pairs with linear optics
\cite{sum1,sum2} or nonlinear optics \cite{sum3}. In 2010, Bonato et
al. \cite{QD2} constructed a complete polarization photonic
Bell-state analyzer with double-sided QD-cavity system in
weak-coupling regime. In 2011, Hu et al. \cite{QD5} also constructed
a complete polarization photonic Bell-state analyzer using both
single-sided and double-sided QD-cavity systems based on GCB. They
pointed out that their schemes work with the coherence photon-spin
interaction in linear regime. Recently, people find that the
spatial-mode entanglement of a photon pair is more robust than its
polarization entanglement. This good feature can be used to design
some robust quantum communication protocols. In this work, we
present a robust two-step QSDC protocol based on the spatial-mode
entanglement and the complete BSA. Certainly, this protocol can be
used to create a private key between two legitimate users, similar
to the works in Refs.\cite{LongLiu,two-step}. The complete BSA of
spatial-mode Bell states in our proposal is also constructed with
the GCB of a one-side QD-cavity system working in linear regime. The
one-side QD-cavity system is used to construct photon-number QND to
distinguish an even number of photons from an odd number of photons
which works as parity check gate in spatial-mode BSA. Here an
external freedom of electron spin in QD is introduced to assist
complete spatial-mode photonic BSA in linear condition.

\begin{figure}[!h]%[tpb]        %Figure 2
\centering\includegraphics[width=8cm,angle=0]{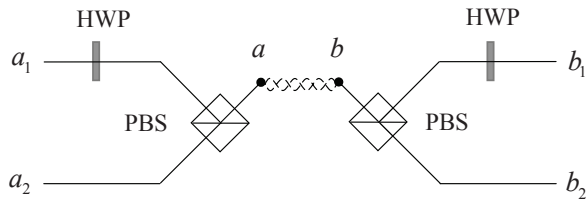}
\caption{Schematic  diagram for the generation of the spatial
entanglement from the polarization entanglement of two-photon
systems.   $a$ and $b$ represent the two photons in a two-photon
system whose initial state is $\frac{1}{\sqrt{2}}(\vert
H\rangle_a\vert H\rangle_b + \vert V\rangle_a\vert V\rangle_b)$. PBS
presents a polarizing beam splitter which is used to transmits the
photon in the horizontal polarization $|H\rangle$ and reflects the
photon in the vertical polarization $|V\rangle$. HWP is a half-wave
plate which is used to flip the polarization states of photons,
i.e., $\vert H\rangle$ $\leftrightharpoons $ $\vert V\rangle$.}
\label{fig3_spatial}
\end{figure}

In experiment, the spatial-mode entanglement of a photon pair can be
obtained by the transformation from the polarization entanglement.
Its principle is shown  in Fig.\ref{fig3_spatial}. That is, one can
first produce a photon pair in an entangled state in the
polarization DOF $\vert
\phi^+_p\rangle_{ab}=\frac{1}{\sqrt{2}}(\vert H\rangle_a\vert
H\rangle_b + \vert V\rangle_a\vert V\rangle_b)$ and then transform
the polarization entanglement into the spatial entanglement with
linear optical elements, shown in  Fig.\ref{fig3_spatial}. When the
photon pair $ab$ passes through the two PBSs and the two HWPs, its
state becomes
\begin{eqnarray}                           \label{eq.add}
\vert \Phi^+\rangle_{ab}=\vert H\rangle_a\vert H\rangle_b\otimes
\frac{1}{\sqrt{2}}(\vert a_1\rangle_a\vert b_1\rangle_b + \vert
a_2\rangle_a\vert b_2\rangle_b).
\end{eqnarray}
This is just the spatial-mode Bell state $\vert \phi^+\rangle_{ab}$.

\begin{figure}                %Figure5
\begin{minipage}[t]{0.5\linewidth}
\centering
\includegraphics[width=1.7in]{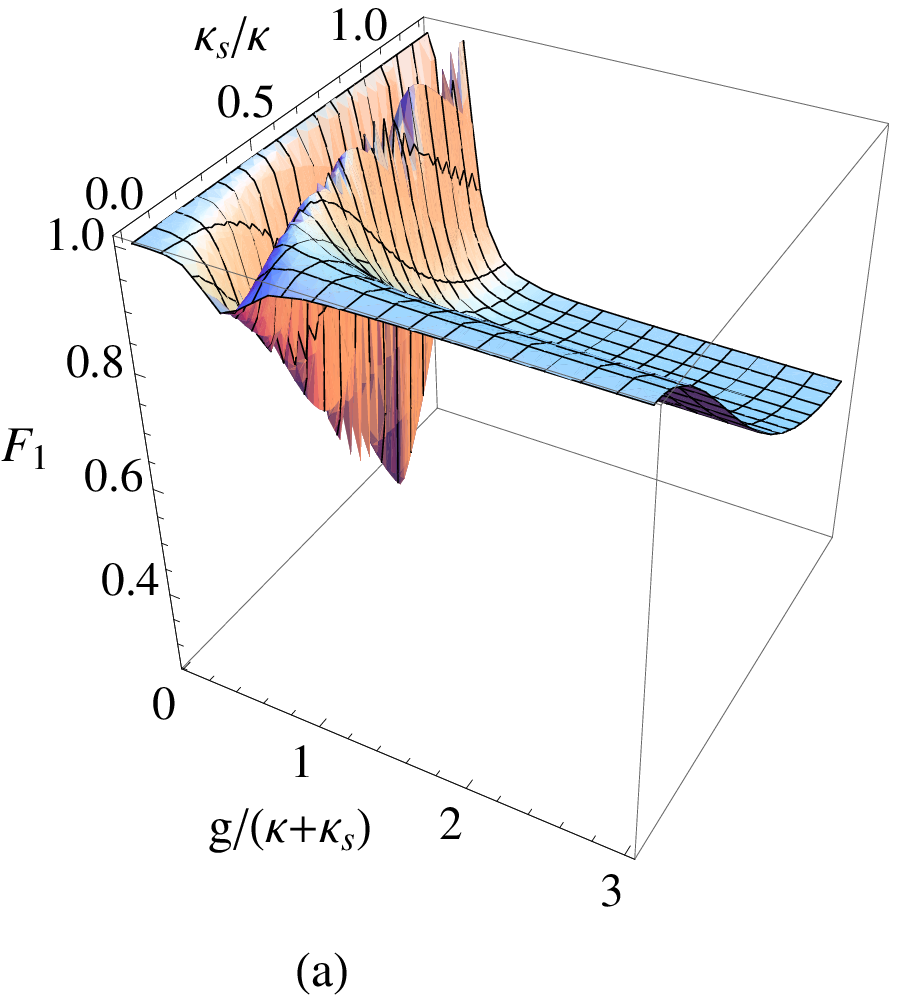}
\end{minipage}%
\begin{minipage}[t]{0.5\linewidth}
\centering
\includegraphics[width=1.7in]{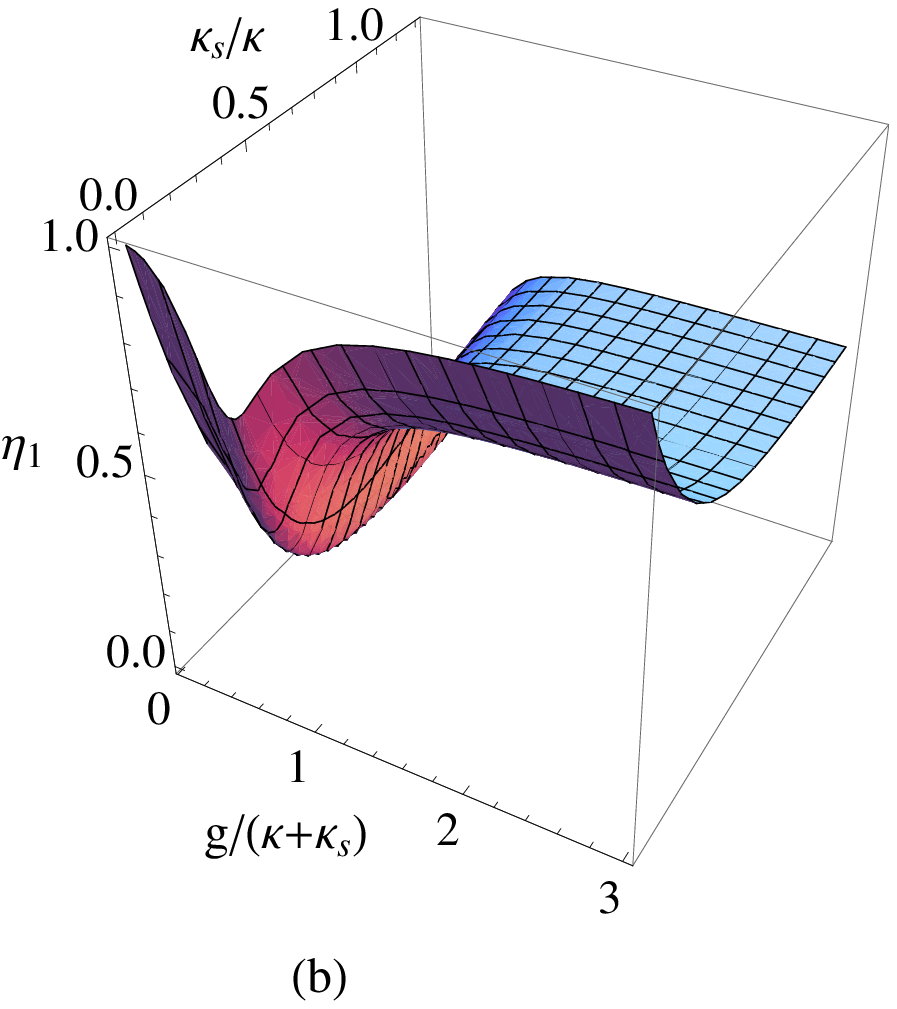}
\end{minipage}
\caption{ The fidelity ($F_1$) and the efficiency ($\eta_1$) of the
present spatial-mode BSA for $|\psi^\pm\rangle$ vs the coupling
strength and different side leakage rates with $\gamma=0.1\kappa$.}
\label{figure3}
\end{figure}

\begin{figure}                %Figure4
\begin{minipage}[t]{0.5\linewidth}
\centering
\includegraphics[width=1.7in]{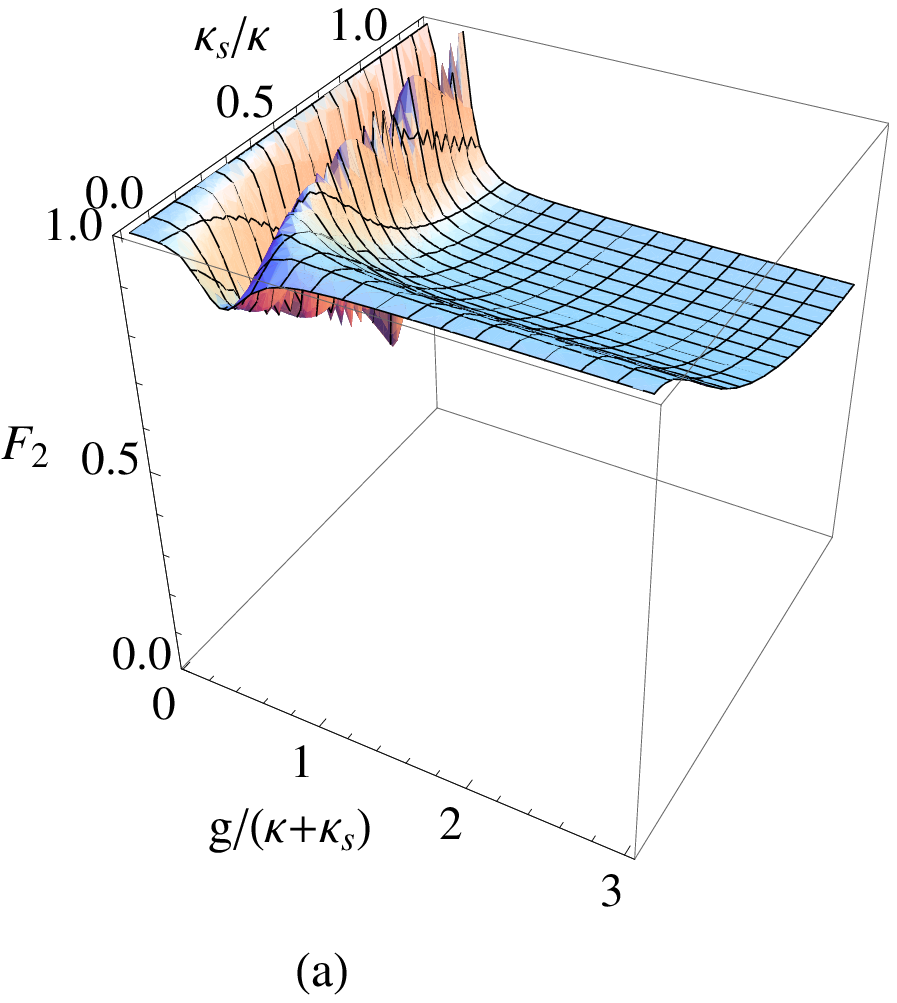}
\end{minipage}%
\begin{minipage}[t]{0.5\linewidth}
\centering
\includegraphics[width=1.7in]{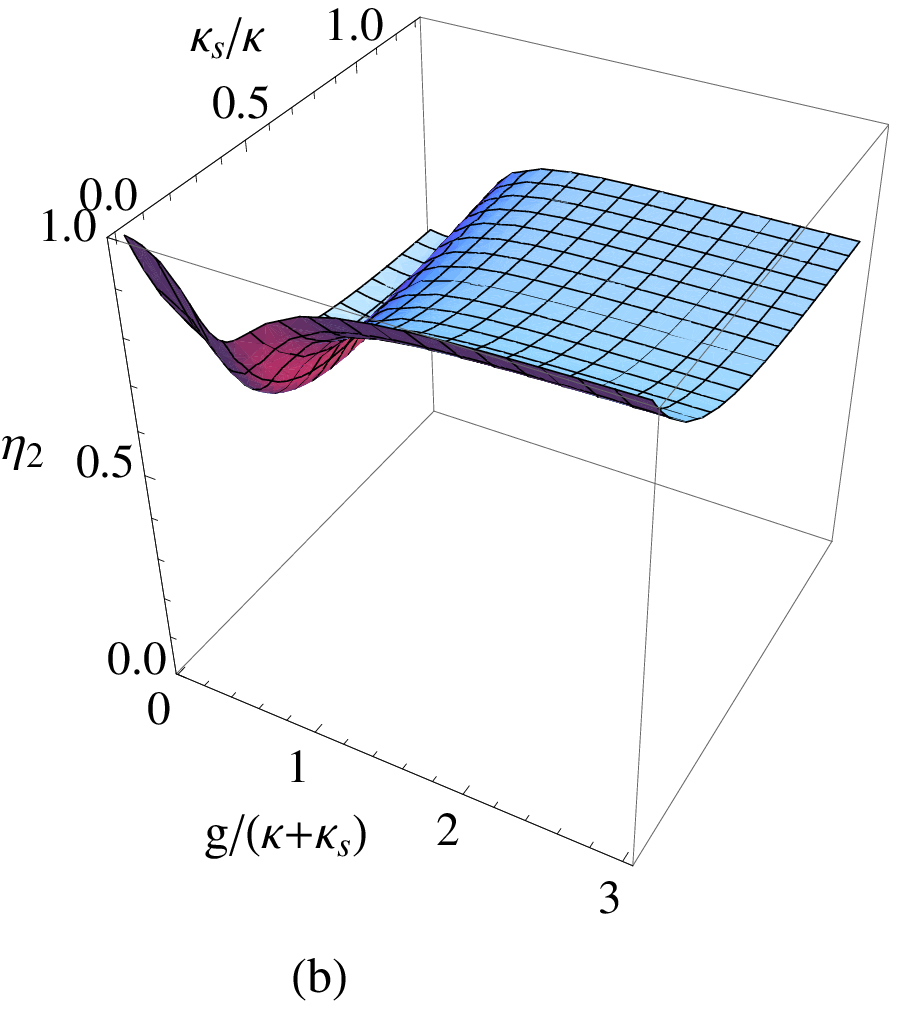}
\end{minipage}
\caption{The fidelity ($F_2$) and the efficiency ($\eta_2$) of the
present spatial-mode BSA for $|\phi^\pm\rangle$ vs the coupling
strength and different side leakage rates with $\gamma=0.1\kappa$.}
\label{figure4}
\end{figure}

In order to construct spatial-mode complete BSA with parity-check
gate, the relative phase shift of the left and the right circular
polarized lights should be $\Delta\varphi=\pm\pi/2$. This can be
achieved by adjusting the frequencies as
$\omega_c=\omega_{X^-}=\omega_0$, $\omega-\omega_c\approx\kappa/2$
and keeping the cavity side leakage rate as $\kappa_s<1.3\kappa$. In
2011, Young et al. \cite{con4} performed an experiment to measure
the quantum-dot-induced phase shift with a quantum dot resonantly
coupled to a pillar microcavity ($Q\sim51000,d=2.5\mu$m), and showed
that a QD-induced phase shift of 0.2 rad between an (effectively)
empty cavity and a cavity with a resonantly coupled QD can be
deduced by using a single-photon level probe and improving mode
matching and pillar design. In an ideal case, the side leakage loss
rate $\kappa_s$ is much lower than the cavity loss rate $\kappa$,
and the reflection coefficients for the cold and the hot cavities
can achieve $|r_0(\omega)|\cong1$ and $|r_h(\omega)|\cong1$ for
unity fidelity and efficiency. However, in experiment, the side
leakage loss of cavity can not be neglected and the fidelity is
reduced to $F=|\langle\psi_f|\psi\rangle|^2$. Here $|\psi_f\rangle$
is the final state of the total system which includes the external
reservoirs and $|\psi\rangle$ is the ideal state without cavity side
leakage. The fidelity and the efficiency of spatial mode BSA process
for states $|\psi^\pm\rangle$ are
\begin{eqnarray}                           \label{eq.11}
F_1 &=& \frac{(|r_0|^3+|r_h|^3+|r_0|^2|r_h|+|r_0||r_h|^2)^2}{4(|r_0|^6+|r_h|^6+|r_0|^4|r_h|^2+|r_0|^2|r_h|^4)},\nonumber\\
\eta_1 &=& \frac{1}{2}|r_0|^4+\frac{1}{2}|r_h|^4,
\end{eqnarray}
and the fidelity and the efficiency of states $|\phi^\pm\rangle$ are
\begin{eqnarray}                           \label{eq.12}
F_2 &=& \frac{(|r_0|^5+|r_h|^5+|r_0|^4|r_h|+|r_0||r_h|^4)^2}{8(|r_0|^{10}+|r_h|^{10}+|r_0|^8|r_h|^2+|r_0|^2|r_h|^8)}\nonumber\\
&&+\frac{(|r_0|+|r_h|)^2}{4(|r_0|^2+|r_h|^2)},\nonumber\\
\eta_2 &=& \frac{1}{2}+(\frac{1}{2}|r_0|^4+\frac{1}{2}|r_h|^4)^2.
\end{eqnarray}
From Fig. \ref{figure3} and Fig. \ref{figure4}, one can see that the
fidelity and the efficiency of the even parity states
$|\phi^\pm\rangle$ are  larger than those of the odd parity states
$|\psi^\pm\rangle$. In Fig. \ref{figure3}(a) and Fig.
\ref{figure4}(a), the fidelity of the present spatial-mode BSA is
relatively high in both the strong coupling regime
($g>(\kappa+\kappa_s)/4$) and the weak coupling regime
($g<(\kappa+\kappa_s)/4$). In Fig. \ref{figure3}(b) and Fig.
\ref{figure4}(b), the high efficiency of the present spatial-mode
BSA can only be gotten in the strong coupling regime. That is, our
BSA proposal  works efficiently in the strong coupling regime, which
is challenging to be observed \cite{couple} in various QD-cavity
systems. The strong coupling strength $g\cong0.5(\kappa+\kappa_s)$
was observed in $d=1.5\,\mu$m micropillar microcavities with the
quality factor $Q=8800$, and this coupling strength can be increased
to be  $g\cong2.4(\kappa+\kappa_s)$ ($Q\sim40000$) \cite{couple1} by
improving the  sample designs, growth, and fabrication
\cite{couple2}. For the odd-parity states $|\psi^\pm\rangle$, if the
coupling strength is $g\cong2.4(\kappa+\kappa_s)$, the fidelity and
the efficiency are $F_1=99.99\%$ and $\eta_1=98.1\%$ with
$\kappa_s/\kappa=0$, and $F_1=69.6\%$ and $\eta_1=53.2\%$ with
$\kappa_s/\kappa=0.7$. Both the fidelity and the efficiency are high
with the strong coupling strength, but they are largely reduced by
the side leakage and cavity loss rate. In experiment, the quality
factor is dominated by the side leakage and cavity loss rate of
micropillar rather than the output coupling rate. In Ref.
\cite{QD5}, Hu et al. reduced the side leakage and cavity loss rate
to $\kappa_s/\kappa\sim0.7$ with the coupling strength
$g\cong\kappa+\kappa_s$ ($Q\sim17000$) by thinning down the top
mirrors of high-Q micropillar ($d=1.5\mu$m). In this case, the
fidelity and the efficiency of states $|\psi^\pm\rangle$ are
$F_1=71.3\%$ and $\eta_1=47.8\%$. The higher the fidelity and the
efficiency are demanded, the smaller $\kappa_s/\kappa$ are required.
This is quite a challenging for micropillar microcavities, because
the strong coupling was achieved in a large micropillar
($d=7.3\mu$m) with a large side leakage \cite{couple3} in recent
experiments.

Besides micropillar microcavities, there are many other factors
which  affect the fidelity of our BSA proposal. The spin decoherence
is the one we should consider with the fidelity decoherence factor
$F'=[1+exp(-\Delta t/T_2^e)]/2$ \cite{QD5}. Here $T_2^e$ is the
electron spin decoherence time and $\Delta t$ is the time interval
between two input photons. As $T_2^e$ could be extended to $\mu$s
with spin echo techniques, which is longer than $\Delta t$ ($n$s) in
the weak excitation approximation, this fidelity decoherence factor
can be neglected. The optical dephasing time of exciton, which is
ten times longer than cavity photon lifetime \cite{trion1, trion2,
trion3},  also reduces the fidelity by a few percent. While the hole
spin in $X^-$ (absence of significant hyperfine interaction) is
three orders longer than the cavity photon life time \cite{trion4,
trion5, trion6}, and the spin dephasing of $X^-$ can be safely
neglected. The hole mixing can also reduce the fidelity due to
unperfect optical selection rule \cite{hole1}. For charged exciton
$X^-$, this effect can be neglected by the quenched exchange
interaction \cite{hole2, hole3}.

In summary, we have presented a complete and deterministic protocol
for the BSA in the spatial-mode DOF of a photon pair, resorting to
the GCB based on a quantum dot inside a one-side micropillar
microcavity. The whole process for BSA is divided into two parts.
First, one divides the four Bell states in spatial-mode DOF into two
groups, that is, the even-parity group and the odd-parity group,
according to their parities of the two photons with a parity-check
QND based on QD-cavity system. Subsequently, one can distinguish the
relative phase of the two states in each group by transforming the
difference of the phases into that of the spatial modes with BSs and
QND. As the spatial entanglement of two-photon systems is more
robust than the polarization entanglement, we presented a robust
two-step quantum secure direct communication protocol with the
spatial-mode Bell states and their BSA. With current technology, our
BSA proposal for spatial-mode Bell states works in both the strong
coupling regime ($g>(\kappa+\kappa_s)/4$) and the weak coupling
regime ($g<(\kappa+\kappa_s)/4$) if the side leakage and cavity loss
rate is small. Maybe this BSA proposal for spatial-mode Bell states
is very useful in the applications in robust quantum communication
protocols, such as QKD, QSS, quantum state sharing, deterministic
secure quantum communication, and so on.

This work is supported by the National Natural Science Foundation of
China under Grant Nos. 10974020 and 11174039,  NCET-11-0031, and the
Fundamental Research Funds for the Central Universities.

\end{document}